\documentclass{ws-ijmpe}

\def\be{\begin{equation}}
\def\ee{\end{equation}}
\def\bea{\begin{eqnarray}}
\def\eea{\end{eqnarray}}
\def\hom{\hbar\omega}
\def\bfr{{\bf r}}
\def\Q{{\hat Q}}
\def\H{{\hat H}}
\def\bfu{{\bf u}}
\def\ur{\bfu(\bfr)}
\def\rhor{\rho(\bfr)}
\def\taur{\tau(\bfr)}
\def\gsb{\langle 0|}
\def\gsk{|0\rangle}
\def\fig#1{Fig.~\ref{#1}}
\def\eq#1{(\ref{#1})}
\def\bgrk#1{\mbox{{\boldmath $#1$ \unboldmath}}\!\!}

\begin{document}

\markboth{Brack, Winkler and Murthy}{Coupling of surface and volume dipole 
                              oscillations in C$_{60}$ molecules}

%
\catchline{}{}{}{}{}
%

\title{COUPLING OF SURFACE AND VOLUME DIPOLE OSCILLATIONS IN
       C$_{60}$ MOLECULES}

\author{\footnotesize M. BRACK} 

\address{Institute of Theoretical Physics, University of Regensburg,
D-93040 Regensburg, Germany\\
e-mail: matthias.brack@physik.uni-regensburg.de}

\author{\footnotesize P. WINKLER} 

\address{Department of Physics, University of Nevada
          Reno, NV 89557, USA}

\author{\footnotesize M. V. N. MURTHY} 

\address{Institute of Mathematical Sciences, CIT Campus,
                Taramani, Chennai 600 113, India}

\maketitle

\begin{history}
\received{(\today)}
\revised{(revised date)}
\end{history}

\begin{abstract}
We first give a short review of the ``local-current approximation'' 
(LCA), derived from a general variation principle, which serves
as a semiclassical description of strongly collective excitations 
in finite fermion systems starting from their quantum-mechanical
mean-field ground state. We illustrate it for the example of 
coupled translational and compressional dipole excitations 
in metal clusters. We then discuss collective electronic dipole 
excitations in C$_{60}$ molecules (Buckminster fullerenes). 
We show that the coupling of the pure translational mode (``surface 
plasmon'') with compressional volume modes in the semiclasscial 
LCA yields semi-quantitative agreement with microscopic 
time-dependent density functional (TDLDA) calculations, while 
both theories yield qualitative agreement with the recent 
experimental observation of a ``volume plasmon''.
\end{abstract}

\section{Introduction}

Early in the history of nuclear physics, the (isovector) giant
dipole resonance (GDR) provided one of the first manifestations
of strongly collective excitations in finite fermion systems. Two 
classical models were suggested to describe the physics of the 
GDR: a) the model of Goldhaber and Teller,\cite{gt} in which
protons and neutrons are both incompressible fluids undergoing
a relative translational oscillation, and b) the model of Steinwedel
and Jensen\cite{sw} (previously also proposed by Migdal\cite{mig}), in 
which protons and neutrons are both locally decompressed or compressed
with opposite phases, such that the total nuclear density remains
constant and a dipole oscillation results (cf.\ \fig{trans} below).
Detailed analysis of
experimental data revealed later that a suitable combination of both 
models was necessary to interpret these data,\cite{mymqj,glei} 
so that the GDR could be classically best understood in terms of 
coupled translational and compressional dipole modes.
These early classical models were later refined by the so-called 
``fluid dynamics''\cite{fludy} and the ``sum rule approach''\cite{blm} 
based on the selfconsistent mean-field description of collective 
excitations in the random phase approximation (RPA).\cite{rowe}
These semiclassical models were successfully used to describe 
collective excitations not only in nuclei, but also in metal 
clusters,\cite{berek,deh93,bra93} where a similar coupling between 
translational and compressional dipole modes has been shown to 
well describe the collective optical response.\cite{bra89,lca,lca3}

In this paper, we review the ``local current approximation''
(LCA), which encompasses both the fluid-dynamical and sum rule 
approaches and can be derived from a variational principle on the same 
footing as the RPA, and quote some of its results for metal clusters. 
We then apply the LCA to collective electronic excitations in 
C$_{60}$ molecules, for which recent experiments\cite{scu1} have 
revealed a ``volume plasmon'', a broad high-energy shoulder in the 
photo-ionization cross section (otherwise dominated by the 
``surface plasmon''\cite{c60be,c60ex}) which again can be 
semiclassically understood as a compressional component of the 
collective dipole excitation.

\section{The local current approximation (LCA)}

The stationary Schr\"odinger equation $\H|\nu\rangle=
(\hat{T}+\hat{V})|\nu\rangle=E_\nu|\nu\rangle$ for a many-body 
system can be cast into the following well-known ``equations of 
motion'':\cite{rowe}
\begin{eqnarray}
\langle 0 | {\cal O_\nu}[\H , {\cal O}_\nu^\dagger ] | 0 \rangle &=&
\hbar \omega_\nu \langle 0 | {\cal O_\nu O_\nu^\dagger} | 0 \rangle,
       \nonumber \\
\langle 0 | {\cal O_\nu}[\H , {\cal O_\nu} ] | 0 \rangle &=& \hbar
\omega_\nu \langle 0 | {\cal O_\nu O_\nu} | 0 \rangle = 0\,,
\label{eom}
\end{eqnarray}
where the operators ${\cal O}_\nu^\dagger$ and ${\cal O_\nu}$ are 
defined by 
\be
{\cal O}_\nu^\dagger |0\rangle=|\nu\rangle\,, \qquad
{\cal O_\nu} |\nu\rangle=|0\rangle\,, \qquad
{\cal O_\nu}|0\rangle=0\,.
\ee 
$|0\rangle$ is the ground state and $\hom_\nu\!=\!E_\nu\!-\!E_0$ 
($\nu\!>\!0$) are the excitation energies. As shown in Ref.\cite{sk}, 
Eq.~\eq{eom} can be rederived by the following variational principle:
\be
\delta E_3[\Q]/\delta \Q=0\,, \qquad
  E_3[\Q]:=\sqrt{m_3[\Q]/m_1[\Q]}\,,
\label{e3}
\ee
where the ``moments'' $m_1$ and $m_3$ -- cf.\ \eq{moms} for their
names -- are defined by
\be
m_1[\Q] := \frac12\langle 0 |\,[\Q,[\H,\Q]]\,|0\rangle,
\quad
m_3[\Q] := \frac12\langle 0 |\,[\,[\H,\Q],
           [[\H,\Q],\H]\,]\,|0\rangle.
\label{m1m3}
\ee
As long as $\Q$ is taken to be the most general (nonlocal) hermitean 
operator, the system (\ref{e3},\ref{m1m3}) is equivalent to the 
exact stationary Schr\"odinger equation. Successive orthogonalization 
of $\Q_1$, $\Q_2$,... yields the exact excitation 
spectrum $E_3(\Q_\nu)=\hom_\nu$ $(\nu=1,2,\dots)$. With
$\Q_\nu\!\propto{\cal O}_\nu^\dagger\!+\!{\cal O}_\nu$, which 
may be interpreted as a set of generalized coordinates (cf.\ 
Refs.\cite{rowe,lca}), we are brought back to \eq{eom}.  

In the
selfconsistent microscopic mean-field approaches, one replaces
$|0\rangle$ either by a Slater determinant (Hartree-Fock theory, HF)
or by the Kohn-Sham (KS) ground state in terms of the local
density $\rho(\bfr)$ (density functional theory, DFT).
If the operator $\Q$ in \eq{m1m3} is replaced by a one-particle-one-hole
($1p1h$) operator, its variation \eq{e3} leads to the RPA equations.
[The RPA ground state should in principle contain $2p2h$ 
excitations, but due to a theorem by Thouless\cite{thou} one may use 
the HF ground state for computing $m_1[Q_\nu]$ and $m_3[Q_\nu]$
in \eq{m1m3}.] In the framework of DFT (using the local density
approximation, LDA, for the exchange-correlation energy), the RPA is 
often also referred to as the time-dependent LDA (TDLDA).

In the sum rule and fluid dynamical approaches, one 
approximates the collective excitation energies $\hom_\nu$
by the energy $E_3[Q_\nu]$, defined as in (\ref{e3},\ref{m1m3}) but 
in terms of suitable local model operators $Q_\nu({\bfr})$, such 
as $Q_d = z$ to describe pure translations (Goldhaber-Teller model), 
the monopole operator $Q_0 = r^2$ for radial compressions 
(``breathing mode''); $Q_2 = r^2Y_{20}$ for quadrupole oscillations, etc.

The {\bf local current approximation (LCA)} consists in the
following assumptions. One uses for $|0\rangle$ in \eq{m1m3} the 
uncorrelated HF or KS ground state, like in RPA or TDLDA, and
takes the operator $\Q$ in (\ref{e3},\ref{m1m3}) to be a {\bf local} 
function $Q(\bfr)$. For local and spin-less (e.g. Coulomb) two-body 
interactions $\hat{V}$, one then obtains
\be
m_1[Q]=m_1[\bfu]=\frac{m}{2\hbar^2}\!\int\!\ur\!\cdot\!\ur\,\rhor\,d^3r\,,
\qquad {\bf u}(\bfr) = -\frac{\hbar^2}{m}\bgrk{\nabla}Q(\bfr)\,,
\label{m1disp}
\ee
where ${\bf u}(\bfr)$ is a local displacement field which is
proportional to the collective current, see \eq{curr} below. 
(This justifies the name of LCA.) Note that if $\nabla^2 Q(\bfr)$
and hence $\bgrk{\nabla}\!\cdot\!\bfu(\bfr)$ is zero, then one has 
incompressible collective flow; otherwise the collective motion 
involves local compression of the fermi fluid.

The moment $m_3[Q]$ is a more complicated functional of $\bfu(\bfr)$, 
of the spatial density $\rhor=\sum_{i=1}^N |\phi_i(\bfr)|^2$ and the 
kinetic energy density $\taur=\sum_{i=1}^N |\nabla\phi_i(\bfr)|^2$ (and 
possibly a current density according to the ``current-DFT''\cite{viko}) 
in terms of the ground-state HF (or KS) wave functions
$\phi_i(\bfr)$. The variation $\delta E_3[Q]/\delta Q(\bfr)=0$ leads 
to fluid dynamical eigenvalue equations:\cite{sk} 
\be
\frac{\delta m_3[\bfu]}{\delta u_j(\bfr)}
= (\hom_\nu)^2\, \frac{m}{\hbar^2}\,\rhor\, u_j(\bfr) \qquad\quad (j=x,y,z)
\label{fdeq}
\ee
yielding the spectrum $\hom_\nu$ and eigenmodes $\bfu_\nu(\bfr)$. 
Eq.~\eq{fdeq} represents three coupled nonlinear fourth-order partial 
differential equations for $u_j(\bfr)$, which in general are 
extremely hard to solve. Because of their dependence on the
wave functions $\phi_i$, one also speaks of ``quantum fluid dynamics''
which includes the effects of zero sound.\cite{fludy}

A practical way (``finite-basis LCA'') to solve \eq{fdeq} 
approximately consists in expanding the operator $Q(\bfr)$ in a 
finite set of basis functions $\{Q_p(\bfr)\}$:
\be
Q(\bfr) = \sum_{p=1}^M c_p Q_p(\bfr)\,.
\label{finbas}
\ee
The variational principle (\ref{e3},\ref{m1m3}) then yields a set 
of $M$ secular equations, whose characteristic equation is:
\be
{\rm det}|C_{pp'}-(\hom_\nu)^2B_{pp'}| = 0 \, ,
         \qquad\quad (p,p',\nu = 1,2,...,M)
\label{seceq}
\ee
with 
\bea
B_{pp'} & = & \gsb \, [Q_p , [H,Q_{p'}]] \, \gsk,\nonumber\\
C_{pp'} & = & \gsb \, [[H,Q_p], [[H,Q_{p'}],H]] \, \gsk.
\label{bc}
\eea
Solution of \eq{seceq} yields the excitation energies $\hom_\nu$ 
and the operators $Q_\nu(\bfr)$ creating the collective states 
$|\nu\rangle$. The characteristic equation \eq{seceq} is the basic
equation of the LCA and looks similar to one version of the RPA 
equations. The basis set $\{Q_p(\bfr)\}$ must be suitably chosen; 
here a good physical intuition for the considered collective 
motion is of great help.

In a further approximation, one may replace the HF or KS ground-state 
density 
by a selfconsistent semiclassical (extended) Thomas-Fermi (ETF)\cite{book}
type smooth density, and use the ETF functional $\tau_{ETF}[\rho]$
(and possibly a corresponding functional for the current density)
in computing the ingredients of \eq{bc}. Although this approximation
misses the quantum shell oscillations in the densities,
it has proven to be sufficient for the evaluation of collective
excitation spectra in many cases.\cite{glei,bra89,lca,lipp}

The LCA is equivalent to the ``generalized scaling model'',\cite{blm} 
representing the system by a collective Hamiltonian
\be
H_{coll}=\frac12\sum_{p,p'=1}^M(B_{pp'}\,{\dot\alpha}_p{\dot\alpha}_{p'}
        +C_{pp'}\,{\alpha}_p{\alpha}_{p'})
\label{hcoll}
\ee
that describes coupled harmonic oscillations with the velocity fields 
${\bf v}_p(\bfr,t)$:
\be
{\bf v}_p(\bfr,t)={\dot\alpha}_p(t)\bfu_p(\bfr)
\ee
and the {\bf local currents} 
\be
{\bf j}_\alpha(\bfr,t)=\rho_{\alpha}(\bfr,t)\,{\bf v}_\alpha(\bfr,t)\,,
\label{curr}
\ee
which obey the continuity equation:
\be
\frac{\partial}{\partial t} \rho_{\alpha}(\bfr,t) +
\nabla\!\cdot {\bf j}_\alpha(\bfr,t) = 0\,.
\ee
Here $\rho_{\alpha}(\bfr,t)$ are the ``scaled'' time-dependent
densities (see Refs.\cite{blm,lca,sk} for their definition).

Having solved either \eq{seceq} or \eq{hcoll}, one knows the
eigenmodes of the system and can calculate its response to an 
external excitation operator {\bf Q}$_{ext}$. To this purpose
one defines a strength function:
\be
S_{{\bf Q}_{ext}}(E)=\sum_{\nu>0}|\langle\nu|{\bf Q}_{ext}|0\rangle|^2
                       \delta(E-\hom_\nu)\,,
\ee
whose energy-weighted moments $m_k({\bf Q}_{ext})$ become:
\be
m_k({\bf Q}_{ext})=\int_0^\infty E^kS_{{\bf Q}_{ext}}(E)\,dE
=\sum_{\nu>0}(\hom_\nu)^k|\langle\nu|{\bf Q}_{ext}|0\rangle|^2.
\label{moms}
\ee
The photo-absorption cross section $\sigma(\omega)$ in the
long-wavelength limit becomes
\be
\sigma(\omega)=(4\pi\omega/3c)\,S_{dip}(E\!=\!\hom)\,,
\ee
where {\bf Q}$_{ext}=Q_{dip}=ez$ is the electric dipole operator 
and $\omega$ is the frequency of the external electric field.

For applications in metal clusters, the following basis set of local
operators\cite{bra89} has successfully been used:
\be
Q_p({\bf r}) = e\,r^p\,Y_{L0}(\theta)\,. \qquad (p=1,2,\dots,M)
\label{modops}
\ee
We consider here only dipole modes ($L=1$); for $p=1$ we then have 
the electric dipole operator $Q_1=Q_{dip}=ez$, while for $p>1$ we 
obtain compressional dipole modes with $\nabla^2 Q_p\neq 0$. Hence, 
a finite set with $M\geq 2$ of these operators will allow for the
description of coupled translational and vibrational dipole modes.
(Alternative basis sets involve spherical Bessel functions.\cite{lca})

\begin{figure}[th]
\centerline{\psfig{file=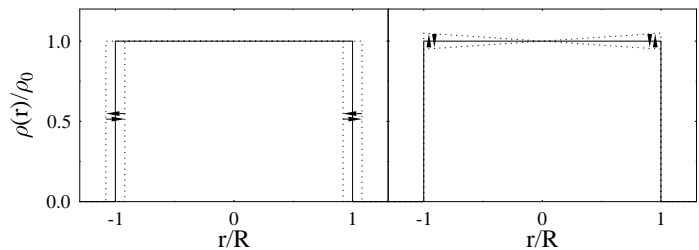,height=3.5cm}}
\caption{Schematic picture of collective oscillations of electrons
(dotted lines) against ions (solid lines) in the spherical jellium
model for a metal cluster. {\it Left:} pure translation, yielding 
the Mie ``surface plasmon'' (cf.\ the Goldhaber-Teller
mode in the nuclear GDR). {\it Right:} compressional dipole mode,
leading to a ``volume plasmon'' (cf.\ the Steinwedel-Jensen-Migdal
mode in nuclei).
}
\label{trans}
\end{figure}

It is illustrative to consider the mechanism of this coupling for
the classical limit of a metal cluster in the spherical jellium 
model.\cite{deh93,bra93,eck} Here the ionic density is taken to be 
a spherical uniform charge distribution with radius $R$:
\be
\rho_I(r) = e\rho_0\,\Theta(R-r)\,,
\label{jelden}
\ee
where the bulk density $\rho_0$ is chosen such that the integrated 
total ionic charge is opposite to that of the valence electrons.
In the classical macroscopic limit $(N\to\infty)$, we take the
electron charge to be opposite to that of the jellium sphere:
$\rho_e(r)=-\rho_I(r)$ and neglect the kinetic and 
exchange-correlation contribution to the total energy, so that 
the system is entirely described by the dominating classical
Coulomb forces. On the left side of \fig{trans}, we illustrate
the pure translational vibration of the electrons against the ions, 
described by the single dipole operator $Q_{dip}=ez$.
It leads to the so-called ``Mie plasmon'' or ``surface plasmon'' 
with the energy
\be
E_3(Q_d) = \hom_{Mie}
         = \sqrt{\hbar^2e^2N/mR^3},
\label{mie}
\ee
where $N$ is the number of single-valenced (e.g. sodium) atoms. 
This is what corresponds to the Goldhaber-Teller model
for the nuclear isovector GDR. On the right side of \fig{trans} 
we sketch a compressional mode brought about by a suitable 
combination of operators \eq{modops} with $p>1$, which 
corresponds to the Steinwedel-Jensen (Migdal) mode for the 
nuclear GDR.

As shown in Ref.\cite{bra89}, the equation \eq{seceq}
can be solved analytically in the above purely Coulombic 
classical limit, with the following interesting solution:\\
{\it When $M>1$ modes are coupled, one of which has $p=1$ 
(pure dipole mode) and all others have $M-1$ different, but
arbitrary real values 
$p>1$, the spectrum always consists of one surface plasmon with 
frequency $\omega=\omega_{Mie}$ and $M-1$ degenerate volume 
plasmons with the frequency $\omega=\omega_{vol}$.}\\ 
This result could only be obtained in Ref.\cite{bra89}
for specific examples. A mathematical proof, for the general 
case of operators \eq{modops} with arbitrary $L$, is given in 
the appendix of this article.

The frequency $\omega_{vol}$ is the bulk plasma 
frequency $\omega_{pl}$ of the corresponding metal
\be
\omega_{vol}=\omega_{pl}=\sqrt{3e^2\!/mr_s^3}\,,
\label{bulkpl}
\ee
$r_s$ being its Wigner-Seitz radius. This ``volume plasmon'' 
can also be brought about as a pure compressional mode 
by the radial (monopole) operator $Q_0=er^2$
(which is also used to describe the nuclear breathing mode):
\be
E_3(Q_0) = \hom_{vol}
         = \sqrt{3\,\hbar^2e^2N/mR^3}
         = \sqrt{3}\,\hom_{Mie}\,.
\label{volpl}
\ee
This mode can, however, not be excited by the external
dipole operator because it corresponds to $L=0$, and
therefore does not couple to the electric dipole field.
The surface plasmon with frequency $\omega_{Mie}$ then
carries all of the dipole strength.

In finite clusters with realistic smooth electronic densities
$\rho_e(r)$, and including the quantum-mechanical kinetic
zero-point and exchange-correlation energies, the operators
\eq{modops} with $p>1$ can couple to the electric dipole
field, the degeneracy of the $M-1$ volume plasmons is broken
and their eigenmodes carry a finite amount of dipole strength.\cite{bra89}
Experimentally, this manifests itself in a broad shoulder,
or sometimes a small extra peak, of the dipole absorption
cross section, located somewhat below the energy of the
bulk plasmon \eq{volpl}. This volume plasmon had already been
anticipated in the early TDLDA calculations of Ekardt.\cite{eck}

\begin{figure}[th]
\begin{center}\vspace{-0.3cm} 
\includegraphics[height=5cm]{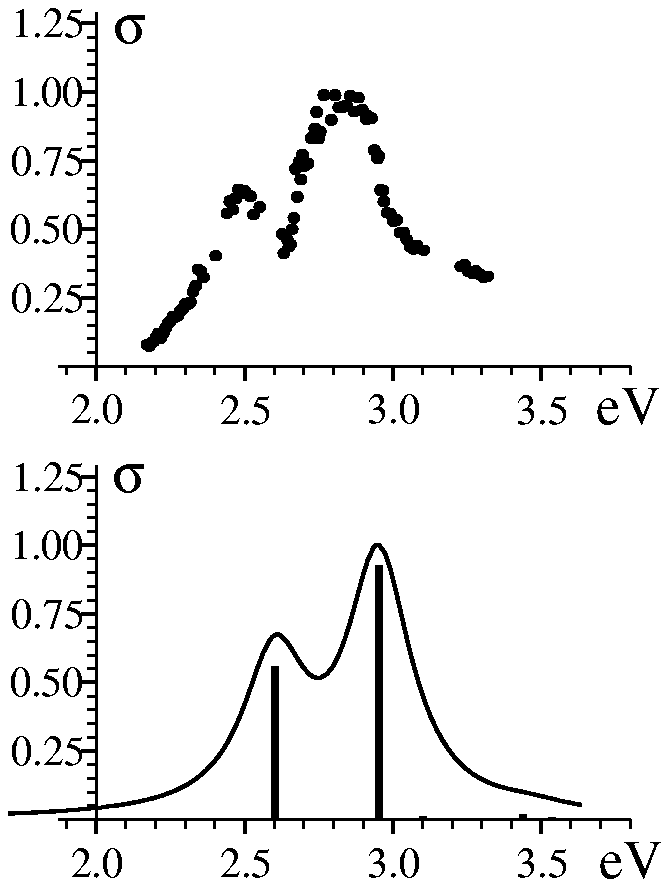}
\includegraphics[height=3.3cm]{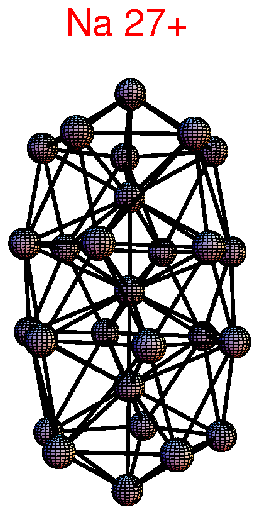}
\hspace{0.4cm}
\includegraphics[height=5cm]{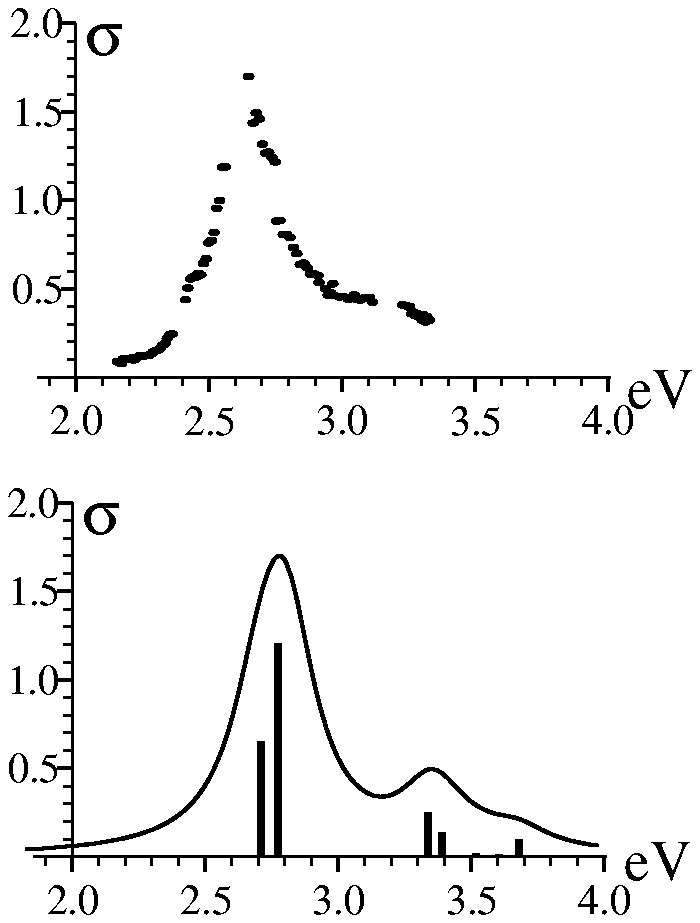}
\includegraphics[height=3.3cm]{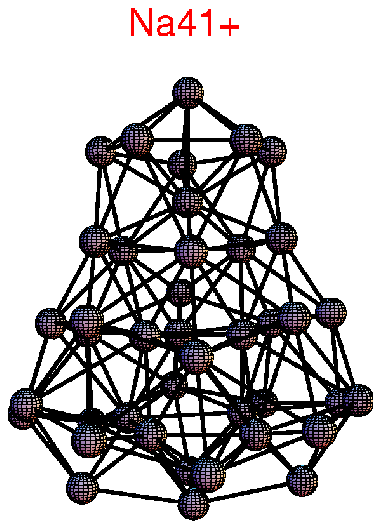}
\end{center}
\caption{Optic dipole response of Na$_{27}^+$ (left part) and 
Na$_{41}^+$ (right part).
{\it Upper left:} Experimental result.$^{24}$
{\it Lower left:} LCA result.$^{14}$ 
{\it Lower right:} Ionic structure from CAPS calculation.$^{14,25}$
}
\label{nacaps}
\end{figure}

In \fig{nacaps} we illustrate the above with some results for
singly-ionized sodium clusters (left side: Na$_{27}^+$, right
side: Na$_{41}^+$). The dots in the upper left part on each 
side give the experimental photo-absorption cross 
section.\cite{naexp2} The lower left
parts give the LCA results;\cite{lca3} the vertical sticks
correspond to the eigenvalues $\hom_\nu$ weighted by
their percentage of the total dipole sum rule, and the
solid curve is obtained by convoluting them with a Lorentzian
to simulate continuum effects (as is customary also in discrete
RPA and TDLDA calculations). On the lower right on each side 
we see the ionic structure,\cite{lca3} as obtained in Car-Parrinello
type molecular calculations in the cylindrically averaged
pseudopotential scheme (CAPS).\cite{caps} We see that the cross 
section of Na$_{27}^+$, with a non-magic number $N=26$ of
valence electrons, clearly exhibits two main peaks which are 
due to the large average deformation of the ionic structure 
(as well-known also for the GDR in deformed nuclei). 
This deformation is prolate, and hence the higher peak has 
roughly twice the strength of the lower peak. These two peaks 
here are both translational surface plasmons, split by the 
average deformation of the system. The volume plasmon is hardly 
recognizable in this case. The cluster Na$_{41}^+$ has a magic 
number\cite{deh93,knig} $N$=40 of valence electrons and 
therefore a nearly spherical electron cloud (in spite
of the non-sphericity of the ionic structure). Here
the response is dominated by one relatively sharp surface peak
and exhibits a plateau in the high-energy shoulder which 
can be identified as a volume plasmon, in fair agreement with
the LCA prediction.
\vspace{-0.8cm}

\section{Coupling of surface and volume modes in C$_{60}$ molecules}

\medskip

\begin{minipage}{10.5cm}
We now turn to the optic response of C$_{60}$ molecules, the
famous Buckminster fullerenes. They consist of 60 carbon atoms,
each providing four valence electrons, so that $N$=240 electrons
can oscillate collectively against the ionic structure. Indeed, a 
giant resonance peak has
\end{minipage}
\begin{minipage}{2.cm}
\hspace{0.1cm}
\includegraphics[height=1.6cm]{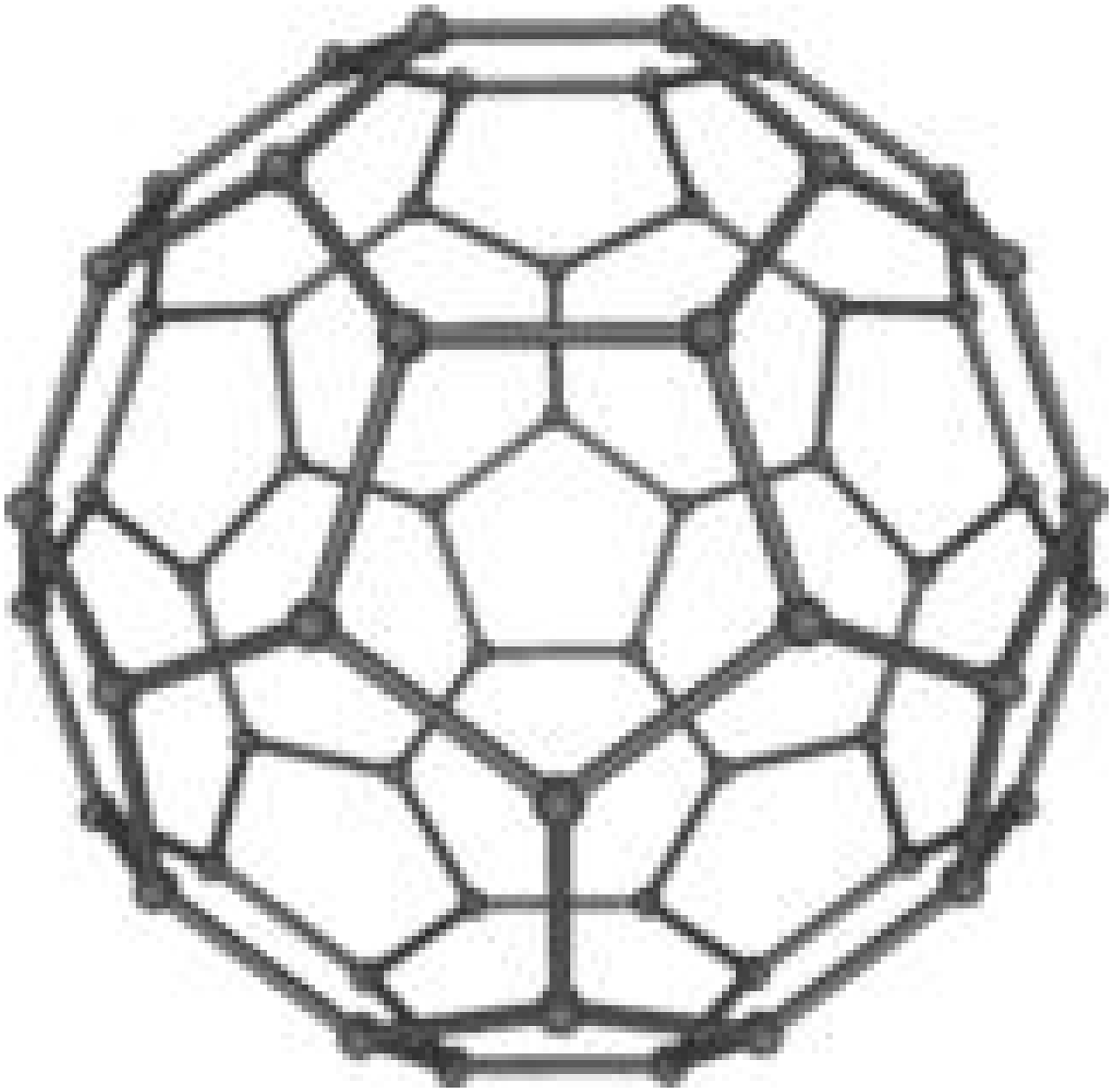}
\end{minipage}\vspace{0.08cm}

\noindent
been predicted\cite{c60be} and experimentally
observed\cite{c60ex} around 20 - 22 eV, and in the framework of the 
jellium model\cite{c60jm} been interpreted as a Mie surface plasmon.

\begin{figure}[th]
\centerline{\psfig{file=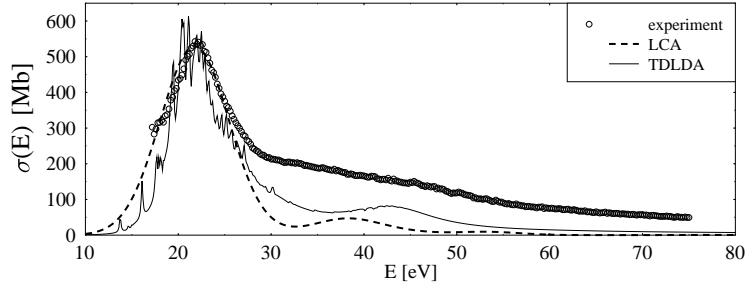,height=4.cm}}
\caption{
Optic response of C$_{60}$ molecules. 
{\it Circles:} Experimental photoionization cross section.$^{15}$ 
{\it Solid lines:} TDLDA calculation$^{15,28}$ and 
{\it dashed lines:} present LCA calculation (see text for details)
of the optic dipole response.
Both theoretical curves have been blue shifted by 5.5 eV
and scaled to the height of the experimental surface peak.
}
\label{results}
\end{figure}
More recent experiments\cite{scu1} on the optic response of C$_{60}$ 
molecules have focused at higher energies, and a broad shoulder was
observed around 30 - 45 eV (see the circles in \fig{results}) which 
was interpreted as a volume plasmon. A two-Lorentzian fit to 
the single photoionization cross section positioned the surface 
plasmon at 22$\pm$0.1 eV and the volume plasmon at 38 $\pm$ 2 eV. 
A TDLDA calculation\cite{scu1,c60td} using a readjusted
version of the jellium model\cite{c60jm} yielded, besides the main
surface peak at 22 eV, an extra peak around 42 eV (see the solid line
in \fig{results}), without however revealing the precise nature of 
the corresponding collective motion. In an ensuing debate, the 
interpretation of the volume peak as a compressional component of the 
collective motion was challenged,\cite{koso} and defended\cite{scu2} 
with reference to the well-known situation in metal clusters.

We shall presently corroborate this interpretation in the LCA. We 
use the same jellium model as in the TDLDA calculation,\cite{scu1}
in which the ionic distribution of the C$_{60}$ molecule is replaced
by a spherical shell with radius $R$ and thickness $\Delta$:
\be
\rho_I(r) = \rho_0\!\left[\Theta\left(R_2-r\right)
           -\Theta\left(r-R_1\right)\right],
\ee
with $R_1=R-\Delta/2$, $R_2=R+\Delta/2$ using the constants
$R=0.354$ nm and $\Delta=0.153$ nm; $\rho_0$ is chosen such that 
the integrated ionic charge is opposite to that of the valence 
electrons.

To obtain a first rough estimate of the results to be expected, 
we use the same schematic classical 
model with $\rho_e(r)=-\rho_I(r)$ and neglecting kinetic and $xc$ 
energies, as discussed above. 
Furthermore we keep only the leading terms in $\Delta/R$ (although
this is not really a small parameter). The energy $E_3(Q)$ with 
the dipole operator $Q_d=ez$ then yields a Mie surface plasmon at
\be
E_3(Q_d) = \hom_{Mie} = \sqrt{\hbar^2Ne^2/3mR^2\Delta}
         = 21.4 \; \hbox{eV}\,, 
\ee
while the monopole operator $Q_0=er^2$ yields a volume plasmon at
\be
E_3(Q_0) = \hom_{vol} = \sqrt{3}\,\hom_{Mie}
         = \sqrt{\hbar^2Ne^2/mR^2\Delta} = 37 \; \hbox{eV}\,, 
\ee
in surprisingly good agreement with the experimentally fitted
peak positions. This is, however, a coincidence, since the volume
plasmon obtained with the monopole operator cannot couple to
the electric dipole field. Coupling $M$ dipole operators of the 
basis set \eq{modops} with $L=1$ and $p=1,2,\dots,M$, we obtain
the following results for the ingredients of the characteristic
equation \eq{seceq}:
\bea
B_{pp'} & = & \frac{\hbar^2}{m}\,\frac{N\,R^{p+p'-2}}{6}\,
              (pp'+2)\,,\label{bppsh}\\
C_{pp'} & = & \left(\frac{\hbar^2}{m}\right)^{\!2}
              \frac{N^2e^2\,R^{p+p'-5}}{3}
              \left\{\frac{R}{2\Delta}\,pp'+\frac16\,(4-p-p'-2pp')\right\}\!.
\label{ingred}
\eea
The equation (\ref{seceq}) can then again be solved analytically 
at leading orders in $\Delta/R$; the calculation is similar to 
that given in the appendix. The resulting spectrum consists of
one surface plasmon at $\hom_{surf}\simeq19.8$ eV and 
$M-1$ degenerate volume plasmons at $\hom_{vol}\simeq31.3$ 
eV. Thus, the coupling of translational and compressional dipole 
modes is seen to shift both peaks towards lower energies, the 
volume peak by a larger amount than the surface peak.

\begin{figure}[th]
\centerline{\psfig{file=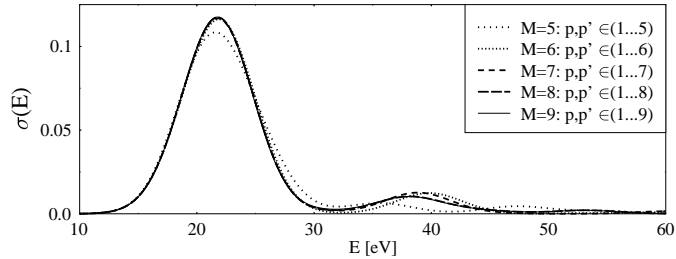,height=4cm}}\vspace{-0.3cm}
\caption{Convergence of the sum-rule weighted dipole response of 
C$_{60}$ molecules in the LCA approximation with respect to the
number $M$ of coupled modes  with $p,p'=1,\dots,M$. The curves 
for $M=8$ (long-dashed) and $M=9$ (solid) cannot be distinguished
(see text for more details).}
\label{lcaconv}
\end{figure}

To be more realistic, we now use the quantum-mechanical KS 
ground-state
density $\rho_e(r)$ of Ref.\cite{c60td}, include the kinetic 
energy using the ETF functional $\tau_{ETF}[\rho]$ (up to 4th 
order, cf. Ref.\cite{book}), and the $xc$ energy in the LDA.
The numerical solutions for the LCA spectrum then converge for 
$M\geq 8$ coupled modes. This is shown in \fig{lcaconv}, where
the sum-rule weighted LCA spectrum has been Lorentzian folded 
with a width of $\Gamma=5$ eV to simulate continuum effects
(which were included in the TDLDA calculation).
\vspace{-0.5cm}
Since the jellium model neglects the ionic structure, the 
collective peaks in both the TDLDA and the LCA calculations
appear at too low energies, as is known to happen also for 
metal clusters.\cite{bra93} Therefore, the TDLDA curve has been 
artificially blue shifted by 5.5 eV in Ref.\cite{scu1}; 
we have done the same with our LCA curves. The converged 
result for $M=8$ corresponds to the dashed curve in \fig{results} 
above. We note that the LCA volume peak now is located at 38 eV, 
exactly like that of the experimental two-Lorentzian fit.

To account for fragmentation channels that were not measured, 
the results of the TDLDA calculations were rescaled\cite{scu1} 
to fit the height of the experimental surface peak; we do the 
same here with our LDA results in \fig{results}. We see that both 
theories underestimate the dipole strength in the high-energy 
region including the volume peak, but give at least a correct 
qualitative result. The LCA result, through the explicit use 
of operators \eq{modops} with $p>1$, confirms the nature of 
the volume peak as due to compressional components of the 
collective electronic motion.

Concerning the difference between the two theoretical results,
for which the same jellium model was used, we note that in the
TDLDA calculation,\cite{scu1,c60td} both $\sigma(n=1)$ and 
$\pi(n=2)$ type valence electrons have explicitly been included 
in the microscopic calculation of the linear respose; the 
$\sigma$ electrons were found to contribute most dominantly in 
the volume peak region. In our LCA calculation we have, however, 
used only one type of valence electrons, which might explain 
the difference of the results particularly in the region of the 
volume peak.
 
\section{Summary and Outlook}

\medskip

We have briefly reviewed the local current approximation (LCA),
a semiclassical approach which can be based on a general variational 
principle on the same footing as the RPA. In the LCA, the ground state 
of a finite fermion system 
is obtained in the selfconsistent mean-field approximation and the
collective excitations are described by coupled local operators 
creating the local currents of the collective motion. This approach, 
which had earlier been successfully used for collective excitations 
in nuclei and metal clusters, has here been applied to the optical 
response of C$_{60}$ molecules. In recent experiments, a volume 
plasmon has been identified in the photoionization cross section 
at an energy $\sim 38\pm 2$ eV, whereas the dominating surface 
plasmon, already earlier known, was located at $\sim 22\pm 0.1$ eV. 
TDLDA calculations reproduce this result qualitatively after 
applying an {\it ad hoc} blue shift to compensate for the missing 
ionic structure in the spherical jellium model used. With the LCA
we obtain very similar results as the TDLDA, using the same jellium
model and applying the same blue shift, and further using a
Lorentzian folding to simulate continuum effects. Since, by explicit 
construction of the coupled local excitation operators, the 
collective currents are known in the LCA, we can identify the 
nature of the volume plasmon in C$_{60}$ as due to compressional
components of the collective electronic motion with respect to
the ions.

It would we worth while to corroborate our semiclassical interpretation
by determining the transition currents both experimentally and
theoretically in microscopic TDLDA calculations. Further improvement
of the theoretical description should include the ionic structure
of the C$_{60}$ molecules. While this might be too time consuming
with purely miscroscopic methods, the LCA appears to be an ideal 
economic tool for this because the semiclassical nature of the 
collective electronic currents is little affected by the ionic 
structure, as known from the correponding results in metal clusters. 

\bigskip

\noindent
{\bf Acknowledgements}

\medskip

\noindent
We are grateful to M. E. Madjet, R. A. Phaneuf, J. M. Rost and 
U. Saalmann for helpful comments and for providing us with
their data used in \fig{results}. One of us (M. B.) 
acknowledges the hospitality of the Physics Department,
University of Nevada at Reno (UNR), during a sabbatical visit, and 
thanks S. K\"ummel and P. G. Reinhard for clarifying discussions.

\bigskip

\noindent
{\bf Appendix: Coupling of surface and volume plasmons in jellium spheres}

\medskip

\noindent
Here we give a proof of the general solution of the characteristic
equation \eq{seceq} surmised in Ref.\cite{bra89} for the classical 
limit with $\rho_e(r)=-\rho_I(r)$ and neglecting kinetic and $xc$ 
energies. We use the set of operators $Q_p(r)=r^pY_{L0}(\theta)$ with 
integer $p=1,2,\dots,M$. The equation then becomes, for fixed angular
momentum $L$,
\be 
\mbox{det}\left | C_{pp'} -(\hom)^2 B_{pp'}\right| 
=f(L)\,\mbox{det}\,|A_{pp'}|=0\,, \qquad (p,p'=1,2,\dots,M)
\label{chareq}
\ee
where $f(L)$ is a factor independent of $p,p'$, and the matrix
$A_{pp'}$ is given by
\be
A_{pp'} = \frac{(2L+1)pp'+L(L+1)(2L-p-p')
          -\lambda\, (2L+1)[pp'+L(L+1)]}{(1+p+p')}\,,
\label{amat}
\ee
and the eigenvalue $\lambda$ is the squared ratio of
frequencies:
\be
\lambda = (\omega/\omega_{vol})^2\,.
\ee
We can rewrite the matrix \eq{amat} in the following form:
\be
A_{pp'} = \frac{(2L+1)[pp'+L(L+1)](1-\lambda) 
-L(L+1)(1+p+p')}{(1+p+p')}\,.
\ee
Now take out a factor $(2L+1)$ from all rows to rewrite 
\eq{chareq} as
\be
\mbox{det}\,|A_{pp'}| = (2L+1)^M\mbox{det}\,|D_{pp'}|=0\,,
\ee
where the matrix $D_{pp'}$ is given by
\be
D_{pp'}=(1-\lambda)\,(pp'+F)/(1+p+p')-G\,,
\ee
and the constants independent of $p,p'$ are defined as
\be
      F=L(L+1)\,, \qquad G=L(L+1)/(2L+1)\,.
\ee

We now consider two cases:

{\bf a) $L=0$:} Then $F=G=0$, and the characteristic equation becomes
\be
\mbox{det}\,|(1-\lambda)\,pp'\!/(1+p+p')|=(1-\lambda)^M 
\mbox{det}\,|pp'\!/(1+p+p')|=0\,.
\ee
Since for any $p,p'>0$ the determinant on the r.h.s. above is never
zero, we get $M$ degenerate solutions with eigenvalue $\lambda=1$, i.e.
with the volume (or bulk) plasma frequency $\omega=\omega_{vol}$ given
in \eq{volpl}.

{\bf b) $L>0$:} In this case $F$ and $G$ are non-zero and the
matrix $D_{pp'}$ has the form
\be
D_{pp'} =
\left(\begin{array}{ccc}
(1-\lambda)(1+F)/3-G \;&\; (1-\lambda)(2+F)/4-G \;&\; ...\\
(1-\lambda)(2+F)/4-G \;&\; (1-\lambda)(4+F)/5-G \;&\; ...\\
(1-\lambda)(3+F)/5-G \;&\; (1-\lambda)(6+F)/6-G \;&\; ...\\
 ...&...& ...
\end{array}\right).
\ee
Notice that in each element, the first term contains the factor 
$(1-\lambda)$ and the second term is the constant $-G$. We now replace 
the first row by the difference between the first and second rows, the 
second by the difference between the second and third, and so on, until 
we reach the last row in which we do not change anything. The determinant, 
whose value is not altered by these manipulations, then becomes:
\be
\mbox{det}\,|D_{pp'}| =
\left|\begin{array}{cccc}
(1-\lambda)\,E_{11} \;&\; (1-\lambda)\,E_{12} \;&\; ... \;&\; (1-\lambda)\,E_{1M} \\
(1-\lambda)\,E_{21} \;&\; (1-\lambda)\,E_{22} \;&\; ... \;&\; (1-\lambda)\,E_{2M} \\
 ... \;&\;  ... \;&\;  ... \;&\;  ... \\
(1-\lambda)\,E_{M1}-G \;&\; (1-\lambda)\,E_{M2}-G \;&\; ... \;&\; (1-\lambda)\,E_{MM}-G 
\end{array}\right|,
\label{detD}
\ee
where the $E_{pp'}$ are linear expressions in the constant $F$. 
Only the in the last row, 

\newpage

\noindent
the additive constant $-G$ remains, 
while all other elements now are proportional to $(1-\lambda)$. 
The characteristic equation therefore becomes
\be
 \mbox{det}\,|D_{pp'}| = (1-\lambda)^{(M-1)} 
                         \mbox{det}\,|{\widetilde E}_{pp'}| =0\,,
\ee
where ${\widetilde E}_{pp'}$ is the remaining matrix after removing
the factor $(1-\lambda)$ from the first $M-1$ rows in \eq{detD}; its 
determinant is linear in $\lambda$. We thus get 
$M-1$ degenerate volume plasmons with eigenvalue $\lambda=1$, i.e. 
with $\omega=\omega_{vol}$ again. The last eigenvalue is difficult 
to find in general. But when any one of the $p$ values 
equals $L$, the last eigenvalue is found to be
$\lambda=LG/F=L/(2L+1)$, corresponding to the Mie plasmon with
angular momentum $L$, i.e., $\omega_L=\sqrt{L/(2L+1)}\,\omega_{vol}$
(cf.\ Ref.\cite{bra89}). 
The generalization of this proof to a set of $M$ arbitrary real 
values of $p$ is straightforward. For the dipole case $L=1$ one 
gets the result stated in the paragraph above \eq{bulkpl}.


\vspace{-0.2cm}

\begin{thebibliography}{99}

\bibitem{gt}    M. Goldhaber and E. Teller, Phys.\ Rev.\ {\bf 74}, 1046 (1948).

\bibitem{sw}    H. von Steinwedel and J. H. D. Jensen, Z. Naturf.\
                Teil A {\bf 5}, 413 (1950). 

\bibitem{mig}   A. B. Migdal, J. Phys.\ USSR {\bf 8}, 331 (1944).

\bibitem{mymqj} W. D. Myers {\it et al.}, Phys.\ Rev.\ C {\bf 15}, 2032 (1977);\\
                J. Meyer, P. Quentin, B. K. Jennings, Nucl.\ Phys.\ A
                {\bf 385}, 269 (1982).

\bibitem{glei}  P. Gleissl, M. Brack, J. Meyer, P. Quentin, Ann.\ Phys.\
                (N.Y.) {\bf 197}, 205 (1990).

\bibitem{fludy} see, e.g., the review by B. K. Jennings and A. D. Jackson,
                Phys.\ Rep.\ {\bf 66}, 141 (1980).

\bibitem{blm}   O. Bohigas, A. M. Lane, J. Martorell, Phys.\ Rep.\
                {\bf 51}, 267 (1979).

\bibitem{rowe}  see, e.g., D. J. Rowe: {\it Nuclear collective
                motion} (Methuen, London, 1970).

\bibitem{berek} G. F. Bertsch and W. Ekardt, Phys.\ Rev.\ B {\bf 32},
                7659 (1985).

\bibitem{bra93} M. Brack, Rev.\ Mod.\ Phys.\ {\bf 65}, 677 (1993).

\bibitem{deh93} W. A. de Heer, Rev.\ Mod.\ Phys.\ {\bf 65}, 611 (1993),

\bibitem{bra89} M. Brack, Phys.\ Rev.\ B {\bf 39}, 3533 (1989).

\bibitem{lca}   P.-G. Reinhard, M. Brack, O. Genzken, Phys.\ Rev.\
                A {\bf 41}, 5568 (1990);\\ 
                P.-G. Reinhard and Y. Gambhir, Ann.\ Phys.\ (Leipzig)
                {\bf 1}, 598 (1992).

\bibitem{lca3}  S. K\"ummel, M. Brack, P.G. Reinhard, Phys.\ Rev.\ B {\bf 58},
                R1774 (1998).

\bibitem{scu1}  S. W. J. Scully {\it et al.}, Phys.\ Rev.\ Lett.\ {\bf 94},
                065503 (2005).

\bibitem{c60be} G. F. Bertsch, A. Bulgac, D. Tom\'anek, Y. Wang,
                Phys. Rev. Lett.\ {\bf 67}, 2690 (1991).

\bibitem{c60ex} I. V. Hertel {\it et al.}, Phys. Rev. Lett.\ {\bf 68}, 784 (1992).
 
\bibitem{sk}    S. K\"ummel and M. Brack, Phys.\ Rev.\ A {\bf 64}, 022506 (2001).
 
\bibitem{thou}  D. J. Thouless, Nucl.\ Phys.\ A {\bf 22}, 78 (1961).

\bibitem{viko}  G. Vignale and W. Kohn, Phys.\ Rev.\ Lett.\ {\bf 77},
                2037 (1996);\\ G. Vignale, C. A. Ullrich, S. Conti,
                Phys.\ Rev.\ Lett.\ {\bf 79}, 4878 (1997).

\bibitem{book}  M. Brack and R. K. Bhaduri: {\it Semiclassical Physics},
                Frontiers in Physics, Vol.\ 96 (revised edition: Westview
                Press, Boulder, 2003); see Ch.\ 4 for the ETF model.

\bibitem{lipp}  E. Lipparini and S. Stringari, Phys.\ Rep.\ {\bf 175}, 103 (1989).

\bibitem{eck}   W. Ekardt, Phys.\ Rev.\ B {\bf 29}, 1558 (1984).

\bibitem{naexp2}M. Schmidt and H. Haberland, Eur.\ Phys.\ J. D {\bf 6}, 109 (1999).

\bibitem{caps}  B. Montag and P.-G. Reinhard, Z. Phys.\ D {\bf 33}, 265 (1995).

\bibitem{knig}  W. Knight {\it et al.}, Phys.\ Rev.\ Lett.\ {\bf 52}, 2141 (1984).

\bibitem{c60jm} M. J. Puska and R. M. Nieminen, Phys. Rev. A {\bf 47}, 1181 (1993). 

\bibitem{c60td} A. R\"udel {\it et al.}, Phys.\ Rev.\ Lett.\ {\bf 89},
                125503 (2002).

\bibitem{koso}  A. V. Korol and A. V. Solov'yov, Phys.\ Rev.\ Lett.\
                {\bf 98}, 179601 (2007).

\bibitem{scu2}  S. W. J. Scully {\it et al.}, Phys.\ Rev.\ Lett.\ {\bf 89},
                179602 (2007).
 
\end{thebibliography}
\end{document}